\documentclass[preprint,fleqn,10pt,prb,balancelastpage,onecolumn]{revtex4}
\usepackage{amsmath}
\usepackage{graphicx}

\input tcilatex

\begin{document}

\preprint{HEP/123-qed}
\title[Spin wave stiffness in band theory ]{The spin angular gradient
approximation in the density functional theory}
\author{M.I.Katsnelson and V.P.Antropov}
\affiliation{Ames Lab, Ames, IA, 50011}

\begin{abstract}
A spin angular gradient approximation for the exchange correlation magnetic
field in the density functional formalism is proposed. The usage of such
corrections leads to a consistent spin dynamical approach beyond the local
approximation. The proposed technique does not contain any approximations
for the form of potential and can be used in modern full potential band
structure methods. The obtained results indicate that the direct 'potential'
exchange in 3d magnets is rather small compared to the indirect 'kinetic'
exchange, thus justifies the dynamical aspect of the local density
approximation in 3d metals.
\end{abstract}

\maketitle

In density functional theory (DFT) two magnetic terms, responsible for any
spin ordering, coexist on both intraatomic and interatomic length scales and
produce a very successful description of static properties in 3d magnets.
One of these terms is the static non-local spin current, which is associated
with the kinetic energy ('kinetic' field (KF)), whereas \ another one is the
large exchange-correlation magnetic field (EXF). While enjoying great
success in weakly correlated materials, the standard \textit{local }spin
density approach (LSDA) for the exchange correlation energy is believed not
appropriate for strongly correlated materials. However, the standard
assumption of the LSDA - a proportionality of EXF and local magnetization $%
\mathbf{m}(\mathbf{r})$ is, in turn appropriate only for strongly correlated
systems (narrow bands), when the magnetic effects related to the 'kinetic'
exchange can be omitted. \ The consequence of this local approximation is
that the large EXF is spin dynamically inactive in the LSDA\cite{PRB1}. This
feature of LSDA restrains the possible range of directions for EXF, which
immediately affects the spin-spin correlation function and, in turn a whole
range of dynamic and thermal properties. Hence, it is highly desirable that
both sources of internal magnetic field are treated with no constraints on
their directions. A general non-local approach should be used to treat
adequately the dynamic aspect of the interrelation between 'kinetic'
(indirect) and 'potential' (direct) magnetic sources. The influence of non
locality of EXF on spin dynamics (SD) of magnets is essentially unknown and
was not studied in real materials. In this paper we undertake an attempt to
overcame this problem and propose a spin angular gradient approximation
(SAGA) for EXF and apply it to the calculations of spin wave (SW) stiffness
in 3d magnetic metals.

The evolution of the many-electron inhomogeneous system in the presence of
the external field is defined in a unique way by the time-dependent
one-particle density matrix $\rho _{\alpha \beta }\left( \mathbf{r,r,}%
t\right) =\left\langle \psi _{\beta }^{+}\left( \mathbf{r},t\right) \psi
_{\alpha }\left( \mathbf{r},t\right) \right\rangle ,$ where $\psi _{\alpha
}\left( \mathbf{r},t\right) $ is the annihilation operator for the electron
at the point $\mathbf{r}$ with spin projection $\alpha $ at the instant time 
$t$. Equivalently, one can introduce the charge $n\left( \mathbf{r,}t\right)
=Tr\rho \left( \mathbf{r,r},t\right) $ and magnetization $\mathbf{m}$ $%
\left( \mathbf{r,}t\right) =Tr\rho \left( \mathbf{r,r},t\right) \mathbf{%
\sigma }$ densities, where $\mathbf{\sigma }$ are Pauli matrices. Starting
from the Schroedinger equation for the many-electron system one can formally
obtain the exact set of equations for these quantities: 
\begin{equation}
\overset{\bullet }{\mathbf{m}}(\mathbf{r,}t)=\gamma \mathbf{m}(\mathbf{r,}%
t)\times \mathbf{B}_{ext}(\mathbf{r,}t)+\frac{i}{2}\mathbf{\nabla }_{\mathbf{%
rr}^{\prime }}^{2}(\rho _{\alpha \beta }\left( \mathbf{r,r}^{\prime }\mathbf{%
,}t\right) \mathbf{\sigma }_{\beta \alpha }-c.c.)_{\mathbf{r}^{\prime }=%
\mathbf{r}},  \label{mexac}
\end{equation}%
\begin{equation}
\overset{\bullet }{n}(\mathbf{r,}t)=\frac{i}{2}\mathbf{\nabla }_{\mathbf{rr}%
^{\prime }}^{2}(\rho _{\alpha \alpha }\left( \mathbf{r,r}^{\prime }\mathbf{,}%
t\right) -c.c.)_{\mathbf{r}^{\prime }=\mathbf{r}}  \label{nexac}
\end{equation}%
where $\mathbf{B}_{ext}(\mathbf{r,}t)$ is the external magnetic field. These
equations are not closed since, generally speaking, the quantity $\mathbf{%
\nabla }_{\mathbf{r}^{\prime }}\rho _{\alpha \beta }\left( \mathbf{r,r}%
^{\prime }\mathbf{,}t\right) _{\mathbf{r}^{\prime }=\mathbf{r}}$ cannot \ be
described directly in terms of $\rho _{\alpha \beta }\left( \mathbf{r,r,}%
t\right) $ (see the discussion of the kinetic energy term in DFT\cite{PZ}).
Introducing the wave functions of Kohn-Sham quasiparticles $\varphi _{\nu
\alpha }$ and making an adiabatic approximation which assumes they evolve
rapidly in comparison to the spin degrees of freedom, we can obtain a 
\textit{closed} set of equations describing charge and spin dynamics: 
\begin{eqnarray}
\overset{\bullet }{\mathbf{m}}(\mathbf{r,}t) &=&\gamma \mathbf{m}(\mathbf{r,}%
t)\times \mathbf{B}_{tot}(\mathbf{r,}t)+  \label{mDF} \\
&&\frac{i}{2}\mathbf{\nabla }_{\mathbf{r}}(\sum_{\nu }^{occ}\varphi _{\nu
\alpha }^{\ast }\left( \mathbf{r,}t\right) \mathbf{\nabla }_{\mathbf{r}%
}\varphi _{\nu \beta }\left( \mathbf{r,}t\right) \cdot \mathbf{\sigma }%
_{\beta \alpha }-c.c.),  \notag
\end{eqnarray}%
\begin{equation}
\overset{\bullet }{n}(\mathbf{r,}t)=\frac{i}{2}\mathbf{\nabla }_{\mathbf{r}%
}(\sum_{\nu }^{occ}\varphi _{\nu \alpha }^{\ast }\left( \mathbf{r,}t\right) 
\mathbf{\nabla }_{\mathbf{r}}\varphi _{\nu \alpha }\left( \mathbf{r,}%
t\right) -c.c.),  \label{nDF}
\end{equation}%
\begin{eqnarray}
&&\left[ \left( -\frac{1}{2}\nabla ^{2}+V_{ext}+V_{H}\right) \delta _{\alpha
\beta }-\mathbf{B}_{ext}\mathbf{\sigma }_{\alpha \beta }+\frac{\delta E_{xc}%
}{\delta \rho _{\beta \alpha }}\right] \varphi _{\nu \beta }\left( \mathbf{r,%
}t\right)   \label{KS} \\
&=&\epsilon _{\nu }\left( t\right) \varphi _{\nu \alpha }\left( \mathbf{r,}%
t\right) ,  \notag
\end{eqnarray}%
where the sum is over the occupied Kohn-Sham states; $V_{ext}$ and $V_{H}$
are the external and Hartree potentials, respectively, $E_{xc}$ is the
exchange-correlation energy, $\mathbf{B}_{tot}=\mathbf{B}_{ext}+\mathbf{B}%
_{xc}$ is the total magnetic field including EXF $\mathbf{B}_{xc}=-\delta
E_{xc}/\delta \mathbf{m.}$ The Kohn-Sham wave functions and the
corresponding energies $\epsilon _{\nu }\left( t\right) $ depend on time due
to the time-dependence of the densities and external field (the latter is
supposed to be weak relative to the characteristic electron energies). 

The second term in Eq.(\ref{mDF}) can be conveniently cast in the following
torque form%
\begin{eqnarray}
\frac{d\mathbf{m}\left( \mathbf{r},t\right) }{dt} &=&\gamma \mathbf{m}(%
\mathbf{r,}t)\times \left[ \mathbf{B}_{kin}(\mathbf{r})+\mathbf{B}_{xc}(%
\mathbf{r})+\mathbf{B}_{ext}(\mathbf{r})\right]   \label{c12} \\
&=&\gamma \mathbf{m}(\mathbf{r,}t)\times \mathbf{B}_{tot}(\mathbf{r})  \notag
\end{eqnarray}%
where the total field acting on the electronic spin at point $\mathbf{r}$ is 
\begin{equation}
\mathbf{B}_{tot}(\mathbf{r})=-\frac{\delta E}{\delta \mathbf{m}(\mathbf{r})}%
=-\frac{\delta T}{\delta \mathbf{m}(\mathbf{r})}-\frac{\delta E_{xc}}{\delta 
\mathbf{m}(\mathbf{r})}+\mathbf{B}_{ext}(\mathbf{r}),  \label{c345}
\end{equation}%
with $\mathbf{B}_{kin}(\mathbf{r})=\mathbf{\nabla }\rho \mathbf{\nabla
\sigma }/m\rho $, $\rho =\sum \phi _{\nu }^{\ast }\phi _{\nu }$, $%
d/dt=\partial /\partial t+\mathbf{v\nabla }$ and $\mathbf{v}$ is a velocity.
From Eq.(\ref{c12}) directly follows that $\mathbf{m}d\mathbf{m}/dt=0.$Below
we use $\mathbf{B}_{ext}=0.$

The basic assumption of LSDA is that $E_{xc}$ is obtained for the
homogeneous electron gas (HEG) model for a given magnetization density $m$,
and assumed to apply to real external potentials. In this local
approximation $\mathbf{B}_{xc}(\mathbf{r})\sim $ $\mathbf{m}(\mathbf{r})$,
hence assuming very strong EXF compared to KF and smallness of dynamics
associated with EXF. Let us now eliminate this requirement and generalize a
shape of EXF for the \textit{spiral} magnetic configuration (spin-density
wave (SDW)) with the amplitude $m$ and the wave vector $\mathbf{Q}.$ Hence,
the term $\mathbf{m}(\mathbf{r,}t)\times \mathbf{B}_{xc}(\mathbf{r})$ will
contribute to Eq.(\ref{c12}).

The SDW state is characterized by anomalous averages $s_{\mathbf{p}%
}=\left\langle c_{\mathbf{p}+\mathbf{Q}/2\uparrow }^{+}c_{\mathbf{p}-\mathbf{%
Q}/2\downarrow }\right\rangle $ where $c_{\mathbf{p}\sigma }^{+}$,$c_{%
\mathbf{p}\sigma }$ are the operators of creation and annihilation of
electrons with momentum $\mathbf{p}$ and spin projection $\sigma .$ Similar
to the Gorkov-Nambu formalism in the theory of superconductivity (see, for
instance, Ref.\cite{KURMAEV}) we introduce the spinor operator $\psi _{%
\mathbf{p}}=(c_{\mathbf{p}+\mathbf{Q}/2\uparrow }^{+},c_{\mathbf{p}-\mathbf{Q%
}/2\downarrow })$. Then the Hamiltonian can be expressed as 
\begin{equation}
H=\sum_{\mathbf{p}}\psi _{\mathbf{p}}h_{\mathbf{p}}\psi _{\mathbf{p}}+\frac{1%
}{2}\sum_{\mathbf{q\neq }0}\sum_{\mathbf{pp}^{\prime }}v_{c}\left( \mathbf{q}%
\right) \left( \psi _{\mathbf{p+q}}^{+}\psi _{\mathbf{p}}\right) \left( \psi
_{\mathbf{p}^{\prime }\mathbf{-q}}^{+}\psi _{\mathbf{p}^{\prime }}\right)
\label{A3}
\end{equation}%
where $v_{c}\left( \mathbf{q}\right) =4\pi e^{2}/\mathbf{q}^{2}V,$ $V$ is a
volume and $h_{\mathbf{p}}=\theta _{\mathbf{p}}+\tau _{\mathbf{p}}\sigma
_{z}-\Delta _{\mathbf{p}}\sigma _{x}.$ Here 
\begin{align}
\theta _{\mathbf{p}}& =\frac{1}{2}\left( \varepsilon _{\mathbf{p}+\mathbf{Q}%
/2}+\varepsilon _{\mathbf{p}-\mathbf{Q}/2}\right) =\mathbf{p}^{2}/2+\mathbf{Q%
}^{2}/8-\mu ,  \label{A5} \\
\tau _{\mathbf{p}}& =\frac{1}{2}\left( \varepsilon _{\mathbf{p}+\mathbf{Q}%
/2}-\varepsilon _{\mathbf{p}-\mathbf{Q}/2}\right) =\mathbf{pQ}/2  \notag
\end{align}%
where $\varepsilon _{\mathbf{p}}=\mathbf{p}^{2}/2-\mu $ is the energy of the
free electron and $2\Delta _{\mathbf{p}}$ is the antiferromagnetic gap,
related to the formation of SDW. In the Hartree-Fock approximation (HFA) the
latter is determined through the relation 
\begin{equation}
\Delta _{\mathbf{p}}=\sum_{\mathbf{p}^{\prime }}v_{c}(\mathbf{p}-\mathbf{p}%
^{\prime })s_{\mathbf{p}^{\prime }}.  \label{A6}
\end{equation}%
Now we will replace $v_{c}$ by the effective Stoner exchange splitting $%
I=\left( V_{exc}^{\uparrow }-V_{exc}^{\downarrow }\right) /\left(
n_{\uparrow }-n_{\downarrow }\right) ,$ where $V_{exc}^{\sigma }=\partial
\left( n\varepsilon _{exc}\right) /\partial n_{\sigma }.$ Then, Eq.\ref{A6}
can be replaced by\ $\Delta =I\left( n_{\uparrow }-n_{\downarrow }\right) /2$%
, where $\Delta $ does not depend on $\mathbf{p}.$

The 'bare' Green function in the Matsubara representation has the form 
\begin{equation}
G\left( i\omega _{m},\mathbf{p}\right) =\frac{1}{i\omega _{m}-h_{\mathbf{p}}}%
=\frac{i\omega _{m}-\theta _{\mathbf{p}}+\tau _{\mathbf{p}}\sigma
_{z}-\Delta _{\mathbf{p}}\sigma _{x}}{\left( i\omega _{m}-\xi _{\mathbf{p}%
_{\uparrow }}\right) \left( i\omega _{m}-\xi _{\mathbf{p}_{\downarrow
}}\right) }  \label{A9}
\end{equation}%
where $\xi _{\mathbf{p}\uparrow ,\downarrow }=\theta _{\mathbf{p}}\mp E_{%
\mathbf{p}}$ is a quasiparticle spectrum for SDW with $E_{\mathbf{p}}=\sqrt{%
\tau _{\mathbf{p}}^{2}+\Delta ^{2}}$. From Eq.\ref{A9} one can find the
occupation number matrix 
\begin{equation}
2N_{\mathbf{p}}=\left( 1+\frac{\tau _{\mathbf{p}}\sigma _{z}-\Delta \sigma
_{x}}{E_{\mathbf{p}}}\right) f_{\mathbf{p}\uparrow }+\left( 1-\frac{\tau _{%
\mathbf{p}}\sigma _{z}-\Delta \sigma _{x}}{E_{\mathbf{p}}}\right) f_{\mathbf{%
p}\downarrow }  \label{A10}
\end{equation}%
where $f_{\mathbf{p}\sigma }=f\left( \xi _{\mathbf{p}\sigma }\right) $ is a
Fermi function. Then for the Fock contribution to the exchange-correlation
energy we will have 
\begin{align}
E_{Fock}& =-\frac{1}{2}\sum_{\mathbf{pp}^{\prime }}v_{c}\left( \mathbf{p}-%
\mathbf{p}^{\prime }\right) Tr\left[ N(\mathbf{p})N(\mathbf{p}^{\prime })%
\right] =E_{Fock}^{(1)}+E_{Fock}^{(2)},  \label{A11} \\
E_{Fock}^{(1)}& =-\frac{1}{4}\sum_{\mathbf{pp}^{\prime }\sigma }v_{c}\left( 
\mathbf{p}-\mathbf{p}^{\prime }\right) f_{\mathbf{p}\sigma }f_{\mathbf{p}%
^{\prime }\sigma }\left( 1+\frac{\tau _{\mathbf{p}}\tau _{\mathbf{p}^{\prime
}}+\Delta ^{2}}{E_{\mathbf{p}}E_{\mathbf{p}^{\prime }}}\right) ,  \notag \\
E_{Fock}^{(2)}& =-\frac{1}{2}\sum_{\mathbf{pp}^{\prime }}v_{c}\left( \mathbf{%
p}-\mathbf{p}^{\prime }\right) f_{\mathbf{p}\uparrow }f_{\mathbf{p}^{\prime
}\downarrow }\left( 1-\frac{\tau _{\mathbf{p}}\tau _{\mathbf{p}^{\prime
}}+\Delta ^{2}}{E_{\mathbf{p}}E_{\mathbf{p}^{\prime }}}\right) .  \notag
\end{align}%
Expression (\ref{A11}) is suitable for the calculation of the exchange
energy in SDW state for any $\mathbf{Q}$ with the corresponding equation for
the chemical potential $\mu $ $\ $

\begin{equation}
N=Tr\sum_{\mathbf{p}}N_{\mathbf{p}}=\sum_{\mathbf{p}}f_{\mathbf{p}\sigma }.
\label{A120}
\end{equation}%
Let us consider an important case of small $\mathbf{Q}$. The expansion of Eq.%
\ref{A120} gives a correction of order $\mathbf{Q}^{2}$ for $\widetilde{\mu }%
=\mu -\mathbf{Q}^{2}/8.$ One has 
\begin{equation}
\delta \widetilde{\mu }=\widetilde{\mu }_{\mathbf{Q}}-\widetilde{\mu }_{%
\mathbf{Q}=0}=-\frac{\mathbf{Q}^{2}}{8F\left( n_{\uparrow },n_{\downarrow
}\right) }  \label{A13}
\end{equation}%
where $F=(p_{F\uparrow }+p_{F\downarrow })I(n_{\uparrow },n_{\downarrow
})/2\pi ^{2}$ is a dimensionless Stoner enhancement factor, $p_{F\sigma
}=(6\pi ^{2}n_{\sigma })^{1/3}.$

Expanding Eq.\ref{A11} up to $\mathbf{Q}^{2}$ we have the following
expression 
\begin{equation}
\frac{E_{Fock}}{V}=-\frac{e^{2}}{8\pi ^{3}}\left\{ \left( p_{F\uparrow
}^{4}+p_{F\downarrow }^{4}\right) -Q^{2}\left[ \left( \frac{1}{2F}-\frac{2}{3%
}\right) \left( p_{F\uparrow }^{2}+p_{F\downarrow }^{2}\right) +\frac{\left(
p_{F\uparrow }+p_{F\downarrow }\right) ^{2}}{12F^{2}}\right] \right\} .
\label{A14}
\end{equation}%
Now let us consider the correlation effects. In the Gell-Mann-Brueckner
theory\cite{GMB} the correlations can be taken into account by using the
polarization operator 
\begin{equation}
\Pi \left( i\omega ,\mathbf{q}\right) =-Tr\sum_{\mathbf{p}%
}T\sum_{\varepsilon _{n}}G\left( \mathbf{p}+\mathbf{q},i\varepsilon
_{n}+i\omega _{n}\right) G\left( \mathbf{p},i\varepsilon _{n}\right) .
\label{A15}
\end{equation}%
With this function the correlation part of $\Omega $-potential can be
expressed as: 
\begin{equation}
\Omega _{corr}=\sum_{\mathbf{q}}\int\limits_{-\infty }^{\infty }\frac{%
d\omega }{4\pi }\left\{ \ln \left[ \frac{1+v_{c}\left( \mathbf{q}\right) \Pi
\left( i\omega ,\mathbf{q}\right) }{1+v_{c}\left( \mathbf{q}\right) \Pi _{%
\mathbf{Q}=0}\left( i\omega ,\mathbf{q}\right) }\right] -v_{c}\left( \mathbf{%
q}\right) \left[ \Pi \left( i\omega ,\mathbf{q}\right) -\Pi _{\mathbf{Q}%
=0}\left( i\omega ,\mathbf{q}\right) \right] \right\} ,  \label{A16}
\end{equation}%
where we wrote only $\mathbf{Q-}$dependent part of the correlation energy.
Substituting Eq.\ref{A9} into Eq.\ref{A15} we have 
\begin{align}
\Pi \left( i\omega ,\mathbf{q}\right) & =\frac{1}{2}\sum_{\mathbf{p}}\left(
1+\frac{\tau _{\mathbf{p}}\tau _{\mathbf{p}+\mathbf{q}}+\Delta ^{2}}{E_{%
\mathbf{p}}E_{\mathbf{p}+\mathbf{q}}}\right) \sum_{\sigma }\frac{f_{\mathbf{p%
}\sigma }-f_{\mathbf{p}+\mathbf{q}\sigma }}{i\omega +\xi _{\mathbf{p}+%
\mathbf{q}\sigma }-\xi _{\mathbf{p}\sigma }}+  \label{A17} \\
& \ 2\left( 1-\frac{\tau _{\mathbf{p}}\tau _{\mathbf{p}+\mathbf{q}}+\Delta
^{2}}{E_{\mathbf{p}}E_{\mathbf{p}+\mathbf{q}}}\right) \frac{f_{\mathbf{p}%
\uparrow }-f_{\mathbf{p}+\mathbf{q}\downarrow }}{i\omega +\xi _{\mathbf{p}+%
\mathbf{q}\downarrow }-\xi _{\mathbf{p}\uparrow }}  \notag
\end{align}%
with 
\begin{equation}
\Pi _{\mathbf{Q}=0}\left( i\omega ,\mathbf{q}\right) =\sum\limits_{\sigma }%
\frac{p_{F\sigma }}{2\pi ^{2}}\left\{ 1+\frac{1}{2q}\left[ p_{F\sigma
}^{2}-\left( \frac{i\omega }{q}+\frac{q}{2}\right) ^{2}\right] \ln \frac{%
i\omega +q^{2}/2+qp_{F\sigma }}{i\omega +q^{2}/2-qp_{F\sigma }}-\frac{1}{2q}%
\left[ p_{F\sigma }^{2}-\left( \frac{i\omega }{q}-\frac{q}{2}\right) ^{2}%
\right] \ln \frac{i\omega -q^{2}/2+qp_{F\sigma }}{i\omega
-q^{2}/2-qp_{F\sigma }}\right\} ,
\end{equation}%
where $\ln $ means the main branch of the logarithm.

Now the corresponding exchange-correlation addition to the spin wave
spectrum at finite $\mathbf{Q}$ can be written as 
\begin{equation}
\delta \omega _{\mathbf{Q}}=\frac{4}{M}\left[ E_{SDW}(\mathbf{Q})-E_{SDW}(0)%
\right] .  \label{A23}
\end{equation}%
where $E_{SDW}(\mathbf{Q})$ is the total energy of the spin spiral and $M$
is a magnetic moment of the cell.

For the electronic gas of high density the main contribution into integral (%
\ref{A16}) comes from the region of small $\mathbf{q}$. At the same time
from Eq.(\ref{A17}) one can see that the interband contribution contains
additional factor $\mathbf{q}^{2}$ compared to the intraband one. Hence, it
seems reasonable to take into account terms of $\mathbf{q}^{2}$ order for
the correlation effects in intraband transitions.

Below we consider non-relativistic case only, when exchange-correlation
energy is invariant with respect to the rotation in spin space. Then, for
the weakly nonuniform distribution of spin density, one can write 
\begin{equation}
E_{exc}=\int d\mathbf{r}\left\{ n\varepsilon _{exc}\left( n_{\uparrow
},n_{\downarrow }\right) +\lambda \left( n_{\uparrow },n_{\downarrow
}\right) D\right\} ,  \label{C45}
\end{equation}%
where $D=\left( \nabla _{\alpha }e_{\beta }\right) \left( \nabla _{\alpha
}e_{\beta }\right) =\left( \nabla \theta \right) ^{2}+\sin ^{2}\theta \left(
\nabla \varphi \right) ^{2}$ is the rotational invariant of lowest order.
Here $\mathbf{e}=\mathbf{m}/\left\vert \mathbf{m}\right\vert \equiv \left(
\sin \theta \cos \varphi ,\sin \theta \sin \varphi ,\cos \theta \right) .$

For the variation of $E_{exc}$ in the spiral structure with $\mathbf{e}(%
\mathbf{r})=\left( \cos \mathbf{Qr},\sin \mathbf{Qr},0\right) $ one can
write $E_{exc}^{Q}-E_{exc}^{Q=0}=V\lambda Q^{2}$. Now, one needs to estimate
the energy of SDW with an amplitude $\left\vert \mathbf{m}\right\vert
=n_{\uparrow }-n_{\downarrow }$ and $\mathbf{Q}\rightarrow 0$. This problem
has been considered by Herring\cite{HER} in HFA. The last one, however,
seems not to be useful in the calculation of the total energy of the SDW due
to an instability at $Q\cong 2k_{F}$\cite{SDW}. At the same time, when
screening is taken into account this instability is destroyed (see, for
instance, Ref.\cite{AMUSIA}). Hence, it is essential to add screening
effects to the Fock contribution to $E_{exc}$. So, the following expression
for the function of $\lambda $ in Eq.\ref{C45} can be obtained%
\begin{equation}
\lambda \left( n_{\uparrow },n_{\downarrow }\right) =-\frac{e^{2}}{16\pi ^{2}%
}\left( \frac{1}{F}-\frac{4}{3}\right) \left( V_{exc}^{\uparrow
}p_{F\uparrow }-V_{exc}^{\downarrow }p_{F\downarrow }\right) +\frac{e^{2}}{%
96\pi ^{3}F^{2}}.  \label{A18}
\end{equation}

This expression properly takes into account the main effect - shift of
chemical potential (\ref{A13}) and exactly corresponds to Eq.\ref{A14} in
the Fock approximation.

Let us discuss the consequences of the addition (\ref{A18}) to
exchange-correlation energy. First of all, EXF $\mathbf{B}_{exc}(\mathbf{r}%
)=-\delta E_{exc}/\delta \mathbf{m}(\mathbf{r})$ is now noncollinear with
respect to the local magnetization, so, the variation of Eq.\ref{C45} leads
to appearance of a new term 
\begin{equation}
B_{exc}^{\gamma }(\mathbf{r})=\frac{2}{m}\left( \delta _{\beta \gamma
}-e_{\beta }e_{\gamma }\right) \nabla _{\alpha }\left( \lambda \nabla
_{\alpha }e^{\beta }(\mathbf{r})\right) .  \label{A20}
\end{equation}

The noncollinearity of $\mathbf{B}_{exc}$ with $\mathbf{e}$ corresponds to
the 'direct' exchange processes, which are absent in LSDA. This field can be
directly included in any traditional full-potential LSDA technique.

Another important consequence of this non-locality is the appearance of an
additional contribution for SWS of ferromagnons 
\begin{equation}
D=\frac{4}{M}\left[ \lim_{\mathbf{Q}\rightarrow 0}\frac{E_{SDW}(\mathbf{Q}%
)-E_{SDW}(0)}{\mathbf{Q}^{2}}\right] .  \label{A21}
\end{equation}
For this contribution we have 
\begin{equation}
\delta D=\frac{4}{M}\int dr\lambda \left( n_{\uparrow },n_{\downarrow
}\right) ,  \label{A22}
\end{equation}%
with integration over the whole elementary cell. This correction represents
an addition to the value of $D$ which was obtained in the random phase
approximation\cite{stiff}.

The numerical calculations were first performed for HEG model. The
calculation of the $\mathbf{Q}$ dependence of the correlation energy
according to Eq.(16) revealed that this contribution is relatively small and
does not have any peculiarities as a function of $\mathbf{Q}$ and has
monotonic behavior. In our approximation it could indicate that this term
contributes to both intraatomic and interatomic range scales in similar
fashion.

Further, we applied the technique described above for the studies of SD in
ferromagnetic (FM) Fe and Ni. These systems are well studied and might be
used as prototype systems for 3d magnetism research. Below we employ a
non-collinear version of the full-potential linear augmented plane wave
method (WIEN code) using both LSDA and generalized gradient correction
approaches. In this approach the magnetization density is treated as a
continuous vector quantity. The parameters of calculations (number of $%
\mathbf{k}$-points, plane waves and etc) were chosen to provide a
convergence of SWS about 3-5\%. Unfortunately, we could not compute the
energy from Eq.(16) for arbitrary $\mathbf{Q}$ with real DFT wave functions
and our study of spin waves in Fe and Ni was performed only for small $%
\mathbf{Q}$, i.e. for SWS from Eq.\ref{A21}. Our standard LSDA calculations
(without SAGA) produced 239 meV/A$^{2}$ and 692 meV/A$^{2}$ for $D$ in Fe
and Ni respectively (see, also Ref.\cite{Mark}). These results are in
agreement with previous theoretical calculations and are close to the
experimental numbers. The non-self-consistent addition of non-local
correction from Eq.\ref{A22} does not change significantly these results,
producing a positive correction of 13 meV/A$^{2}$ and 45 meV/A$^{2}$ for Fe
and Ni respectively (so, the total $D$ become 253 meV/A$^{2}$ and 735 meV/A$%
^{2}$). \ The addition of Eq.\ref{C45} for the self-consistency procedure
and the total energy calculations modifies these results. The
self-consistent account of SAGA increases the total $D$ up to 271 meV/A$^{2},
$ with the increasing EXF contribution of 21 meV/A$^{2}$. The effect of the
increase of the total $D$ was also found in Ni where EXF contributes 94 meV/A%
$^{2}$ to the total self-consistent value of 782 meV/A$^{2}.$ This result
already indicates that SD in 3d metals exists mostly due to 'kinetic'
(indirect) exchange with relatively small contribution from 'potential'
(direct) exchange. Whereas the absolute value of the latter term is rather
large, its non-local (dynamic) part is much weaker compared to 'kinetic'
spin current dynamics. Overall the addition of SAGA leads to an increase of
the total $D$ from both the direct SAGA addition and the indirect
modification of 'kinetic' contribution (effects of self-consistency).
Whereas this result is obtained only for small $\mathbf{Q,}$ we expect that
at larger $\mathbf{Q}$ this effect will be more pronounced. Also one can
argue that the addition of diagrams beyond the GW approximation can be
important due to their influence on Stoner exchange splitting. Both LSDA and
GW calculations produce much larger splitting than the one observed on
experiment, hence one can expect that the other diagrams will correct it and
as a result will generate a smaller SWS.

In summary, we estimated the non local part of the exchange correlation
magnetic field in real magnets. Our results for Fe and Ni indicated that in
spite of the large absolute value of the exchange correlation magnetic
field, the non-local part of this field, which contributes to microscopical
spin dynamics is rather small compared to the 'kinetic' spin current term in
3d ferromagnetic metals. We expect that this correction can be important in
the interstitials of weak magnets, systems with strong interaction between
the local moment and conduction electrons, and magnets with non-collinear
ordering.

This work was carried out at the Ames Laboratory, which is operated for the
U.S.Department of Energy by Iowa State University under Contract No.
W-7405-82. This work was supported by the Director for Energy Research,
Office of Basic Energy Sciences of the U.S.Department of Energy.

\end{document}